\documentclass[letterpaper, 10 pt,conference]{ieeeconf}

\IEEEoverridecommandlockouts                              % This command is only needed if 
\usepackage{graphicx}
\usepackage{mathptmx} % assumes new font selection scheme installed
\usepackage{times} % assumes new font selection scheme installed
\usepackage{amsmath} % assumes amsmath package installed
\usepackage{amssymb}  % assumes amsmath package installed
\usepackage{cite}
\usepackage{algorithm}
\usepackage{algorithmic}
\usepackage{subfig}
\usepackage{float}%提供float浮动环境
\usepackage{booktabs}%提供命令\toprule、\midrule、\bottomrule

\title{\LARGE \bf
EDRF: Enhanced Driving Risk Field Based on Multimodal Trajectory Prediction and Its Applications
}

\author{Junkai Jiang$^{1}$, Zeyu Han$^{1}$, Yuning Wang$^{1}$, Mengchi Cai$^{1}$, Qingwen Meng$^{1}$, Qing Xu$^{1}$ and Jianqiang Wang$^{1\ast}$
\thanks{This research was funded by National Natural Science Foundation of China (Grant No. 52131201), National Natural Science Foundation of China under Grant 52302499, the China Postdoctoral Science Foundation under Grant 2023TQ0171, the Postdoctoral Fellowship Program of CPSF under grant GZC20231298, and the Shuimu Tsinghua Scholarship of Tsinghua University. This research was also supported by the Tsinghua University-Didi Joint Research Center for Future Mobility.}% <-this % stops a space
\thanks{$^{1}$Junkai Jiang, Zeyu Han, Yuning Wang, Mengchi Cai, Qingwen Meng, Qing Xu and Jianqiang Wang are with the School of Vehicle and Mobility, Tsinghua University, Beijing, China}
\thanks{$^{\ast}$Corresponding author: Jianqiang~Wang (wjqlws@tsinghua.edu.cn).}
}

\begin{document}

\maketitle
\thispagestyle{empty}
\pagestyle{empty}

\begin{abstract}

Driving risk assessment is crucial for both autonomous vehicles and human-driven vehicles. The driving risk can be quantified as the product of the probability that an event (such as collision) will occur and the consequence of that event. However, the probability of events occurring is often difficult to predict due to the uncertainty of drivers’ or vehicles’ behavior. Traditional methods generally employ kinematic-based approaches to predict the future trajectories of entities, which often yield unrealistic prediction results. In this paper, the Enhanced Driving Risk Field ($EDRF$) model is proposed, integrating deep learning-based multimodal trajectory prediction results with Gaussian distribution models to quantitatively capture the uncertainty of traffic entities’ behavior. The applications of the $EDRF$ are also proposed. It is applied across various tasks (traffic risk monitoring, ego-vehicle risk analysis, and motion and trajectory planning) through the defined concept Interaction Risk ($IR$). Adequate example scenarios are provided for each application to illustrate the effectiveness of the model.
\end{abstract}

\section{Introduction} \label{Sec:Intro}

The emerging autonomous driving (AD) technology has developed rapidly in recent years, for its potential to solve problems in traditional traffic, such as driving safety and traffic congestion \cite{zhou2022developing}. To achieve AD, it is necessary to integrate multiple advanced technologies. Generally, there are four key components in the framework of AD, including environmental perception, situation awareness (SA), decision making, and vehicle control \cite{li2020threat}. Among these modules, SA analyses the outputs of the environmental perception, and provides the inputs such as safety assessment for decision making. Thus, it serves as a bridge between perception and decision making, which is of crucial significance.

In \cite{endsley1995toward}, SA is defined as three stages: perception of the environment, comprehension of the situation, and projection of the future states. SA in AD technology also follows this definition, including environmental understanding, risk assessment, motion and trajectory prediction, etc. However, the hierarchical relationship between them is relatively vague. According to different research focuses, \cite{wang2024enabling} takes advantage of the results of trajectory prediction for risk assessment; \cite{wang2024pregsu} views trajectory prediction as a downstream task of environmental understanding to verify its effectiveness; \cite{wang2023vif} encodes risk assessment as features and input them into the trajectory prediction network. Therefore, the combination of different tasks in SA has basically become a consensus. In this paper, we mainly focus on risk assessment with the consideration of environmental modeling and multimodal trajectory prediction, to improve traffic safety for both autonomous vehicles (AVs) and human-driven vehicles (HDVs).

The traffic risk can be quantified as the product of the probability that an event (such as collision) will occur and the consequence of that event \cite{naatanen1976road}. However, the probabilities of events occurring are often difficult to predict due to the presence of human drivers. In recent years, deep learning technologies have been extensively applied in the field of motion and trajectory prediction of vehicles \cite{zhou2023query, zhou2023qcnext, lan2023sept}. By modeling the interactions between vehicles, as well as between vehicles and environment, it provides new insights for predicting future trajectories and offers innovative methods for quantifying the probabilities of events in risk assessment. The outcomes of trajectory prediction are typically one or more trajectories with their respective probabilities, which to some extent capture the uncertainty associated with drivers’ behavior. However, this is still insufficient. Further effort is needed to address the challenges of quantifying the uncertainty inherent in drivers’ behavior for accurate risk assessment.

Furthermore, the application of risk assessment methods remains an area that requires deeper exploration. Existing approaches are often scenario-driven, which can be adequate for simple scenarios but tend to underperform in complex situations. Additionally, the smooth transition between scenarios presents a challenge \cite{kolekar2020human}. Therefore, there is a need to explore unified methods for applying risk assessment across varying contexts.

therefore, this paper mainly focuses on the traffic risk assessment and its applications. Fig~\ref{fig:framework} illustrates the core framework of this paper, which mainly includes the modeling of the enhanced driving risk field ($EDRF$) and the application areas. Our contributions and innovative points include:

  \begin{figure*}[htbp]
      \centering
      \includegraphics[width=16cm]{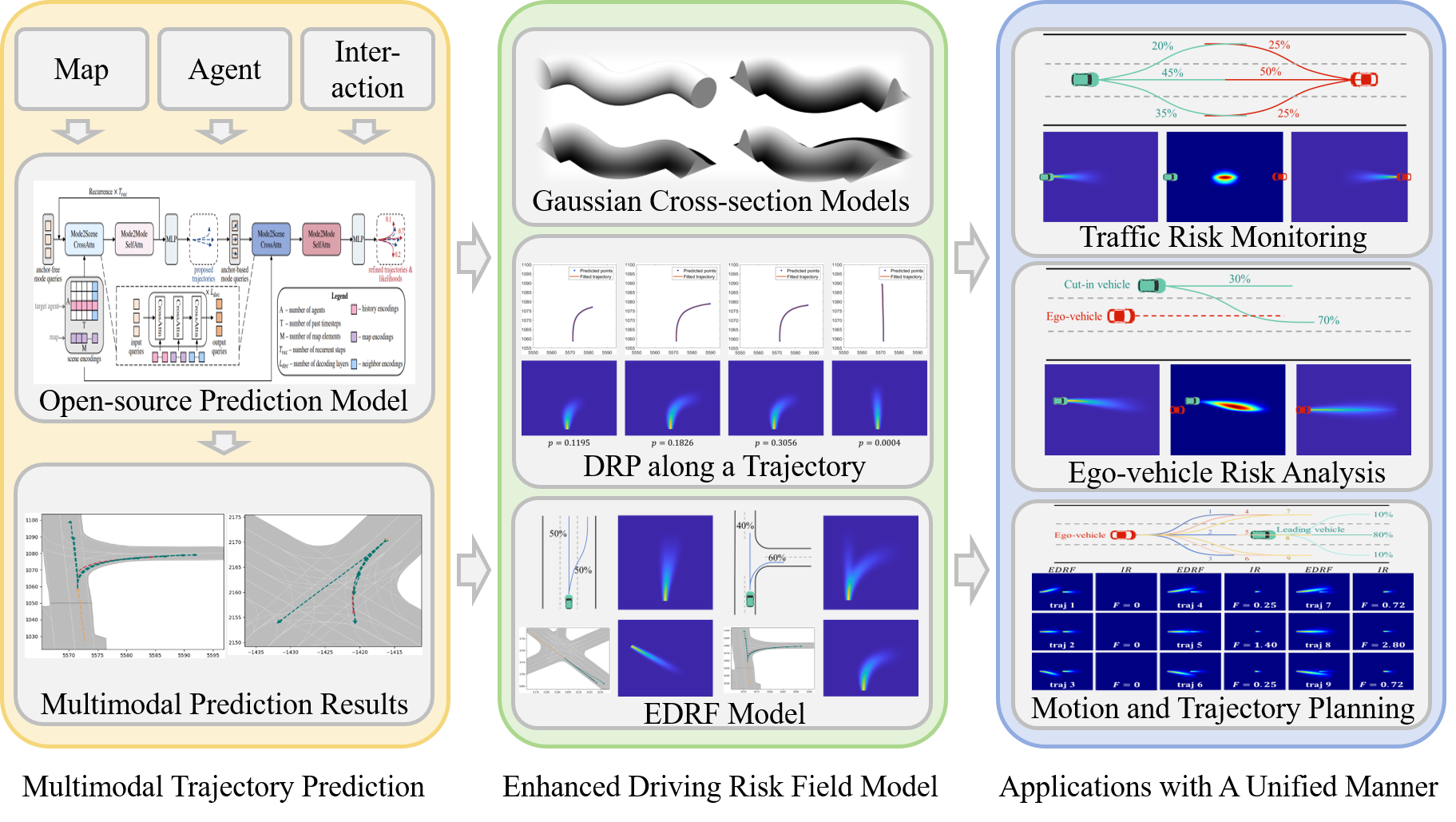}
      \vspace{-5pt}
      \caption{The framework of Enhanced Driving Risk Field ($EDRF$) and its applications.}
      \vspace{-10pt}
      \label{fig:framework}
  \end{figure*}

\begin{itemize}
    \item{By integrating multimodal trajectory prediction with Gaussian distribution models, we quantitatively capture the uncertainty of drivers’ behavior. }
    \item{Combined with the field model, we propose the $EDRF$ which is aware of drivers’ uncertainty.}
    \item{The $EDRF$ is applied across various tasks using a unified manner, with application examples provided to validate the effectiveness of the model.}
\end{itemize}

The remainder of this paper is organized as follows. Section \ref{Sec:RelatedWorks} provides a concise review of related work. Sections \ref{Sec:EDRF} describes the $EDRF$ in details and Section \ref{Sec:Applications} introduces its applications in different fields of risk assessment with examples. Finally, section \ref{Sec:Conclusion} draws the conclusions.

\section{Related Works} \label{Sec:RelatedWorks}
Trajectory prediction is crucial for understanding the environment, whether for AVs or HDVs, and it significantly influences the decision making of the vehicles or drivers. Existing trajectory prediction methods can generally be categorized into traditional model-based approaches and the emerging deep learning-based approaches. The former primarily relies on the current state of the vehicle and pre-established models to output a predicted trajectory for a specified future period. Typical methods include those based on vehicle kinematics \cite{berthelot2011handling}, Gaussian process models \cite{gao2020situational}, and Monte Carlo simulations \cite{althoff2011comparison}. Limited by the expressive capability of the models, the performance of these methods cannot be guaranteed, and they struggle to quantify the uncertainty of the predicted trajectories in uncertain traffic environments. On the contrary, deep learning-based methods often yield multimodal trajectory predictions, providing more valuable information for vehicle decision-making and planning. Unlike unimodal prediction that forecasts a single future path, multimodal trajectory prediction estimates several potential trajectories. The recent availability of public trajectory prediction datasets, such as Argoverse \cite{chang2019argoverse}, have greatly facilitated the advancement of these methods. Several prediction methods \cite{zhou2023qcnext, lan2023sept} have achieved promising outcomes on public datasets, with most of the metrics being satisfactory. The focus of this paper is not on exploring prediction methods, but rather on risk assessment based on multimodal prediction results. Hence, we employ the latest open-source prediction algorithms from the academic community as the foundation for our subsequent risk assessment module.

Risk assessment has consistently been a focal point in the fields of AD and advanced driving assistance system (ADAS), owing to the critical importance of safety when driving on the road. In \cite{li2020threat}, key metrics for risk assessment are categorized into five types, i.e., time-based metrics (time headway, time to collision, time to reaction) \cite{vogel2003comparison, hayward1972near}, kinematic-based metrics (minimum distance, required acceleration) \cite{spiess1997highway}, statistics-based metrics (collision probability, machine learning methods) \cite{jansson2005collision, wang2015driving1,wu2023towards}, potential field-based metrics \cite{wang2015driving2, reichardt1994collision}, and unexpected driving behavior-based metrics (unusual driving events, traffic conflicts) \cite{worrall2010improving}. Among these methods, potential field-based ones can uniformly model the risks posed by all traffic elements in complex environments, resulting in the broadest range of applicability. For instance, \cite{wang2015driving2} defines a driving safety field composed of potential field, kinetic field and behavioral field, which comprehensively quantify the risks formed by the driver, vehicle, and road. However, the modeling of the behavioral field is insufficient and struggle to fully capture the uncertainty of vehicles behavior. Therefore, this paper will integrate multimodal trajectory prediction and Gaussian models with the potential field model, to develop the $EDRF$ which not only retains the unified advantages of the field model but also enhances the model’s ability to express the uncertainty associated with vehicle behavior.

Applications of risk assessment can be divided from two perspectives: that of traffic management and that of ego-vehicle. From the perspective of traffic management, risk assessment can be employed for traffic risks monitoring, risk situations analysis in hazardous scenarios, and traffic conflict analysis. Conversely, from the perspective of ego-vehicle, risk assessment is utilized in ADAS, including rear-end collision warning system \cite{hayward1972near}, automatic emergency braking system \cite{spiess1997highway}, lane keeping system \cite{reichardt1994collision}, adaptive cruise control \cite{vogel2003comparison}, and active collision avoidance system \cite{hayward1972near}. Additionally, risk assessment also plays a crucial role in the decision-making module of AD, which encompasses motion and trajectory planning \cite{worrall2010improving,deng2023deep}. As previously mentioned, the application of current risk assessment methods is often scenario-driven, with different methods or metrics applied under different scenarios. We aim to address this situation by fully leveraging the advantages of the field model, applying the $EDRF$ across various tasks in a unified manner, thereby enhancing the generalizability of the method.

\section{Model of the Enhanced Driving Risk Field} \label{Sec:EDRF}

In this section, we provide a detailed introduction to the modeling process of the $EDRF$. Initially, we utilize the open-source model QCNet \cite{zhou2023query} to output results of multimodal trajectory prediction. Building on this, we present the driving risk probability ($DRP$) model along a predicted trajectory, incorporating Gaussian distribution models. Finally, the complete $EDRF$ model is proposed.

\subsection{Multimodal Trajectory Prediction}

The model is trained with Argoverse 2 dataset. The prediction precision of models used in this paper and the state-of-the-art (SOTA) are shown in Table~\ref{table1}, where Final Displacement Error (FDE) and Average Displacement Error (ADE) are chosen as the metrics. 

% TODO:TABLE1
\begin{table}[htbp]
    \renewcommand\arraystretch{1.3}
    \centering
	\caption{Precision of Trajectory Prediction Models}
        % \vspace{-5pt}
	\label{table1}
	\begin{center}
		\begin{tabular}{ccc}
			\toprule  
                $\mathbf{Model}$ & $\mathbf{ADE}$\ (m) & $\mathbf{FDE}$\ (m) \\
			\hline
                QCNet (open-source) \cite{zhou2023query} & 0.62 & 1.19 \\
                QCNeXt \cite{zhou2023qcnext} & 0.50 & 1.02 \\
                SEPT \cite{lan2023sept} & 0.61 & 1.15 \\
			\bottomrule
		\end{tabular}
	\end{center}
\end{table}
% \vspace{-5pt}

\begin{figure}[htbp]
  \centering
  % 第一行两张图片
  \begin{minipage}{0.48\linewidth}
    \centering
    \includegraphics[width=0.9\linewidth]{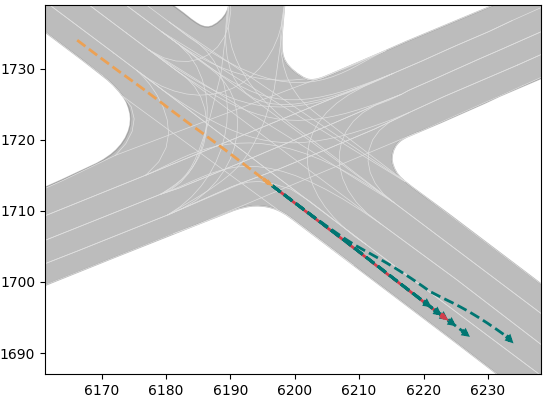} 
    \centerline{(a)}
  \end{minipage}
  \begin{minipage}{0.48\linewidth}
    \centering
    \includegraphics[width=0.9\linewidth]{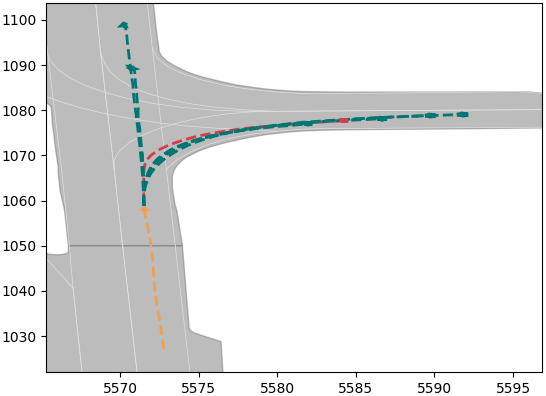}
    \centerline{(b)}
  \end{minipage}
  \begin{minipage}{0.48\linewidth}
    \centering
    \includegraphics[width=0.9\linewidth]{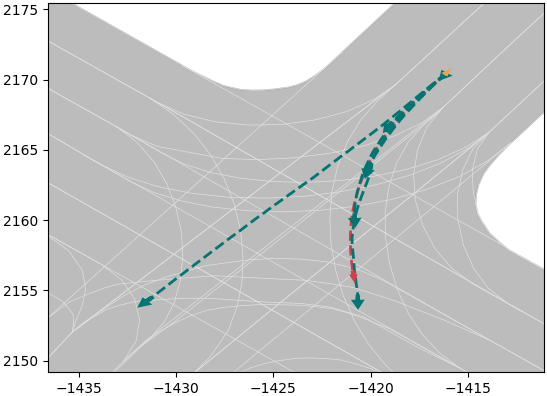}
    \centerline{(c)}
  \end{minipage}
  \begin{minipage}{0.48\linewidth}
    \centering
    \includegraphics[width=0.9\linewidth]{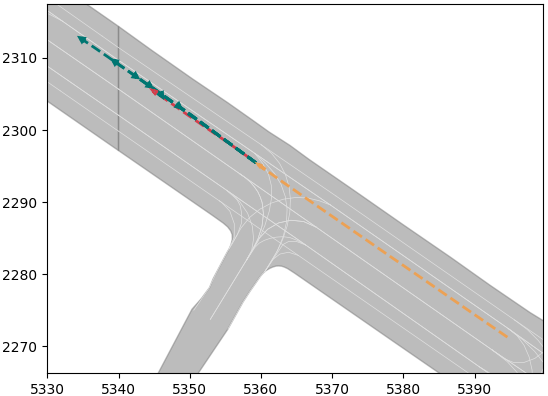}
    \centerline{(d)}
  \end{minipage}
  \caption{Examples of the multimodal trajectory prediction results.The orange line represents the vehicle’s historical trajectory, the red line denotes the ground truth, and the green lines depict the multimodal trajectory prediction results.}
  \label{fig:traj_pred_exams}
\end{figure}

It is worth noting that we have not made efforts to improve the accuracy of trajectory prediction for several reasons. Firstly, the gap between results of model used in this paper and the SOTA is already minimal. Secondly, trajectory prediction inherently possesses a high degree of uncertainty, and our focus is on modeling this uncertainty rather than enhancing the precision of deterministic outcomes. Thirdly and most importantly, trajectory prediction serves merely as an independent input module within our method. Should more precise results be required in the future, a simple model replacement could be readily implemented. Examples of the trajectory prediction results are illustrated in Fig.~\ref{fig:traj_pred_exams}, where the orange line represents the vehicle's historical trajectory, the red line denotes the ground truth, and the green lines depict the multimodal trajectory prediction results.

% todo:fig2

% \vspace{-15pt}
\subsection{$DRP$ along a Predicted Trajectory}

\begin{figure}[htbp]
  \centering
  % 第一行两张图片
  \begin{minipage}{1\linewidth}
    \centering
    \includegraphics[width=0.95\linewidth]{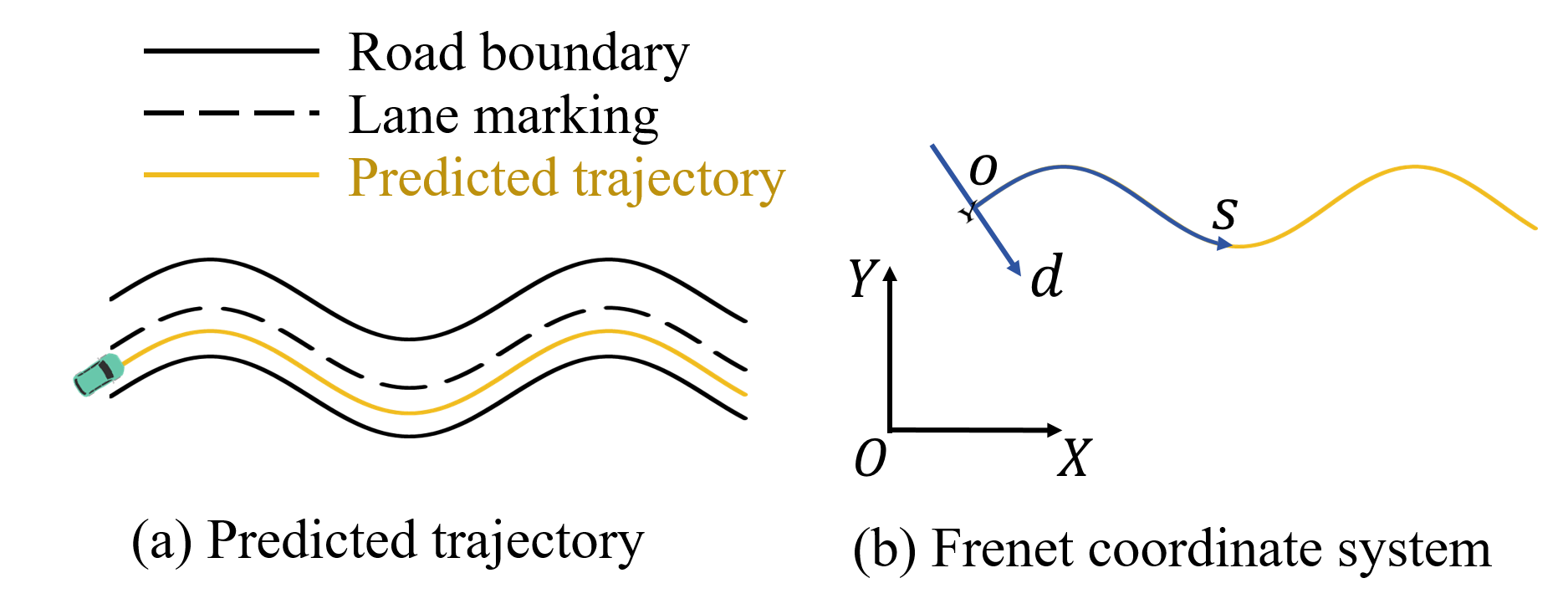} 
  \end{minipage}
  % \vspace{-5pt}
  \caption{The predicted trajectory and the Frenet coordinate system built upon it.}
  \label{fig:frenet}
\end{figure}

\begin{figure}[t]
  \centering
  \begin{minipage}{1\linewidth}
    \centering
    \includegraphics[width=0.9\linewidth]{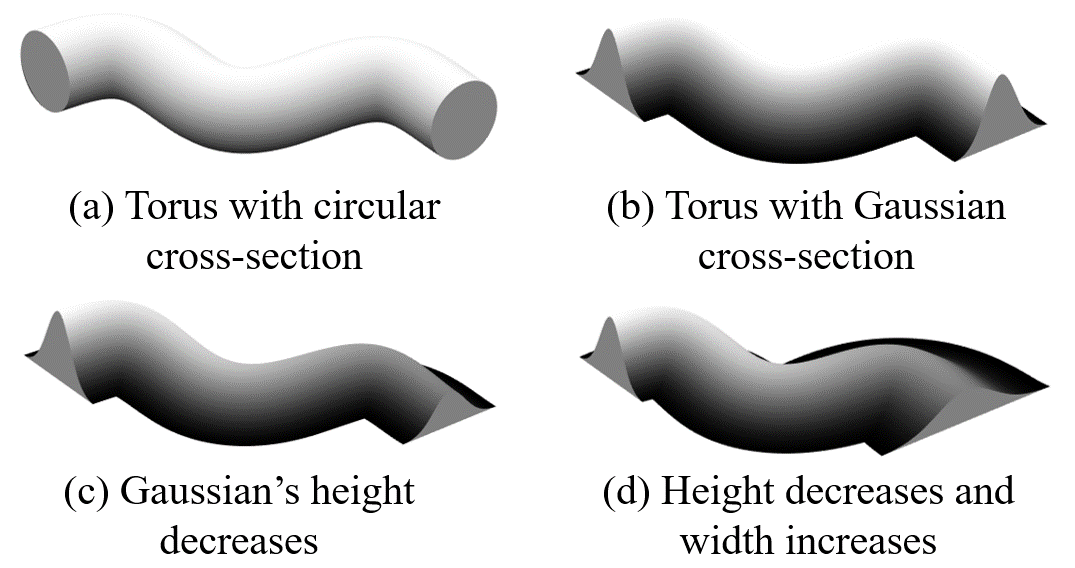} 
  \end{minipage}
  \vspace{-5pt}
  \caption{The modelling process of the $DRP$. We finally obtain the torus with Gaussian cross-section, whose height decreases and width increases along the predicted trajectory.}
  \label{fig:torus}
\end{figure}

In \cite{kolekar2020human}, the authors proposed the driver’s risk field and modelled it as a torus with a Gaussian cross-section along a straight line or an arc, which is determined by the current vehicle’s steering angle and velocity. We expand the model by extending its future trajectory from a kinematic-based straight line or arc to the predicted trajectory. With the current vehicle position as the origin, a Frenet coordinate system $sod$ is established along the predicted trajectory (Fig.~\ref{fig:frenet}). The height and width of the Gaussian cross-section are functions of the predicted trajectory length, denoted as $s$ (Fig.~\ref{fig:torus}).

% \vspace{-5pt}

\begin{equation}
DRP(x,y)=DRP(s,d)=a(s) \cdot \exp{(-\frac{d^2}{2\sigma(s)^2})}
\end{equation}

Here, $(s,d)$ represents the corresponding coordinates of point $(x,y)$ when transferred from the Cartesian coordinate system to the Frenet coordinate system. The parameters $a$ and $\sigma$ are the height and width of the Gaussian distribution respectively, and they are defined as follows:

\begin{equation}
a(s)=q(s-s_{pt})^2
\end{equation}

\begin{equation}
\sigma(s)=(b+k\cdot \overline{\kappa_{pt}})s+c
\label{eqn:sigma}
\end{equation}

$s_{pt}$ denotes the total length of the predicted trajectory. The height $a(s)$ is modelled as a parabola, based on the assumption that the risk decreases relatively rapidly around the vehicle \cite{wang2015driving2}, and approaches zero at the end of the trajectory. Parameter $q$ represents the steepness of the parabola. The width parameter $\sigma$ is modeled as a linear function of the path length $s$, with the slope determined by the parameters $b$ and $k$, as well as the average curvature $\overline{\kappa_{pt}}$ of points along the trajectory. The parameter $c$ represents the initial Gaussian width. This modeling approach implies that the width of the Gaussian distribution increases along the trajectory, reflecting increasing uncertainty in the predictions, which aligns with intuitive understanding. Additionally, incorporating the $k\cdot \overline{\kappa_{pt}}$ term into the slope of the linear model suggests that higher average curvatures lead to more pronounced uncertainty dispersion, consistent with the greater uncertainty observed in turning trajectories in reality.

\begin{figure}[htbp]
  \centering
  \begin{minipage}{1\linewidth}
    \centering
    \includegraphics[width=0.95\linewidth]{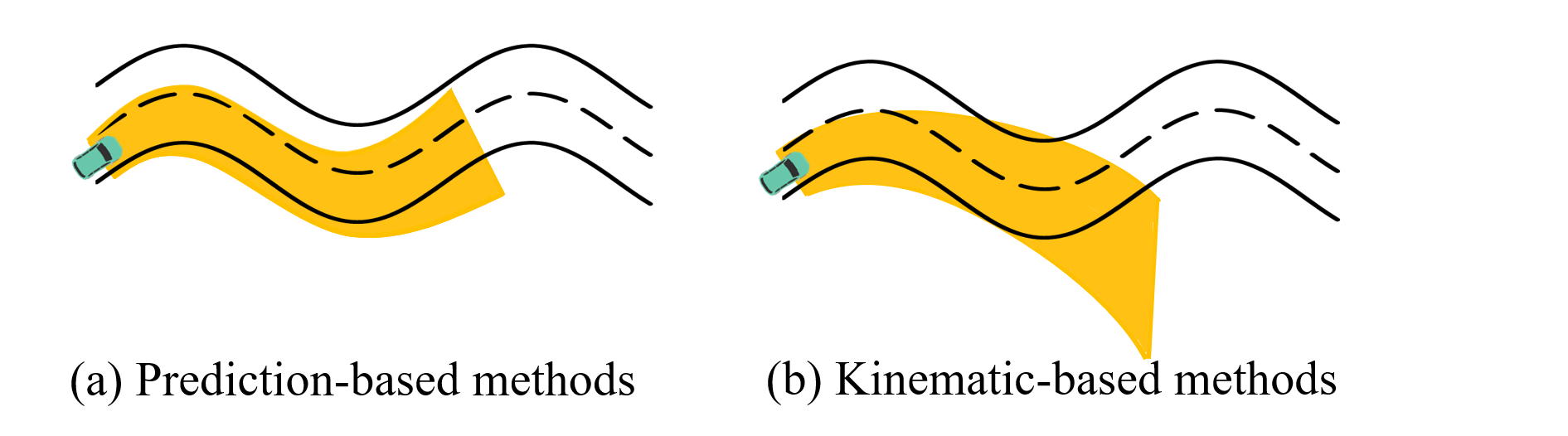} 
  \end{minipage}
  \caption{Comparison of deep learning prediction-based and kinematic-based future trajectory prediction methods.}
  \label{fig:prediction-good}
\end{figure}

\begin{figure}[htbp]
  \centering
  \begin{minipage}{1\linewidth}
    \centering
    \includegraphics[width=1\linewidth]{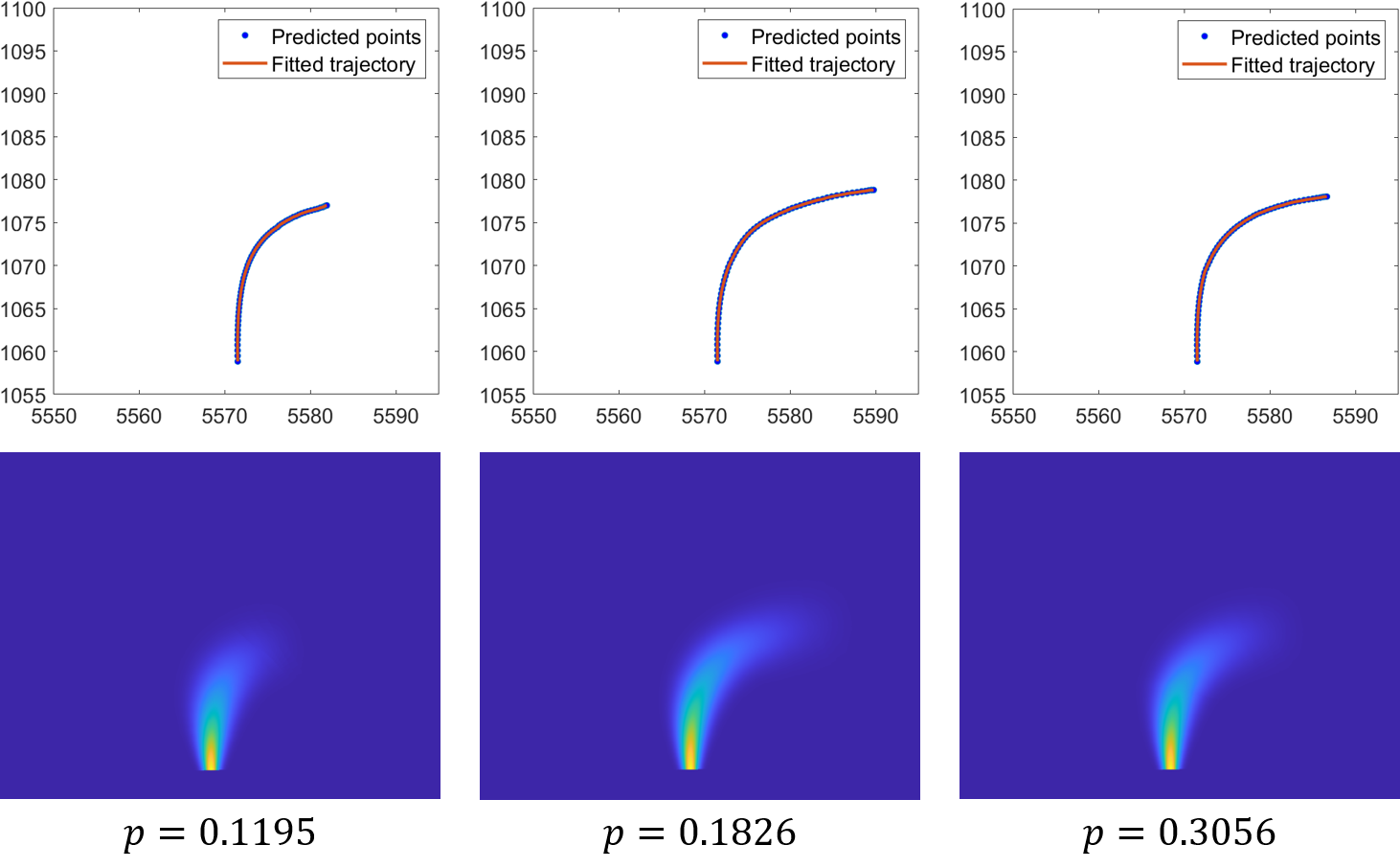} 
  \end{minipage}
  \begin{minipage}{1\linewidth}
    \centering
    \includegraphics[width=1\linewidth]{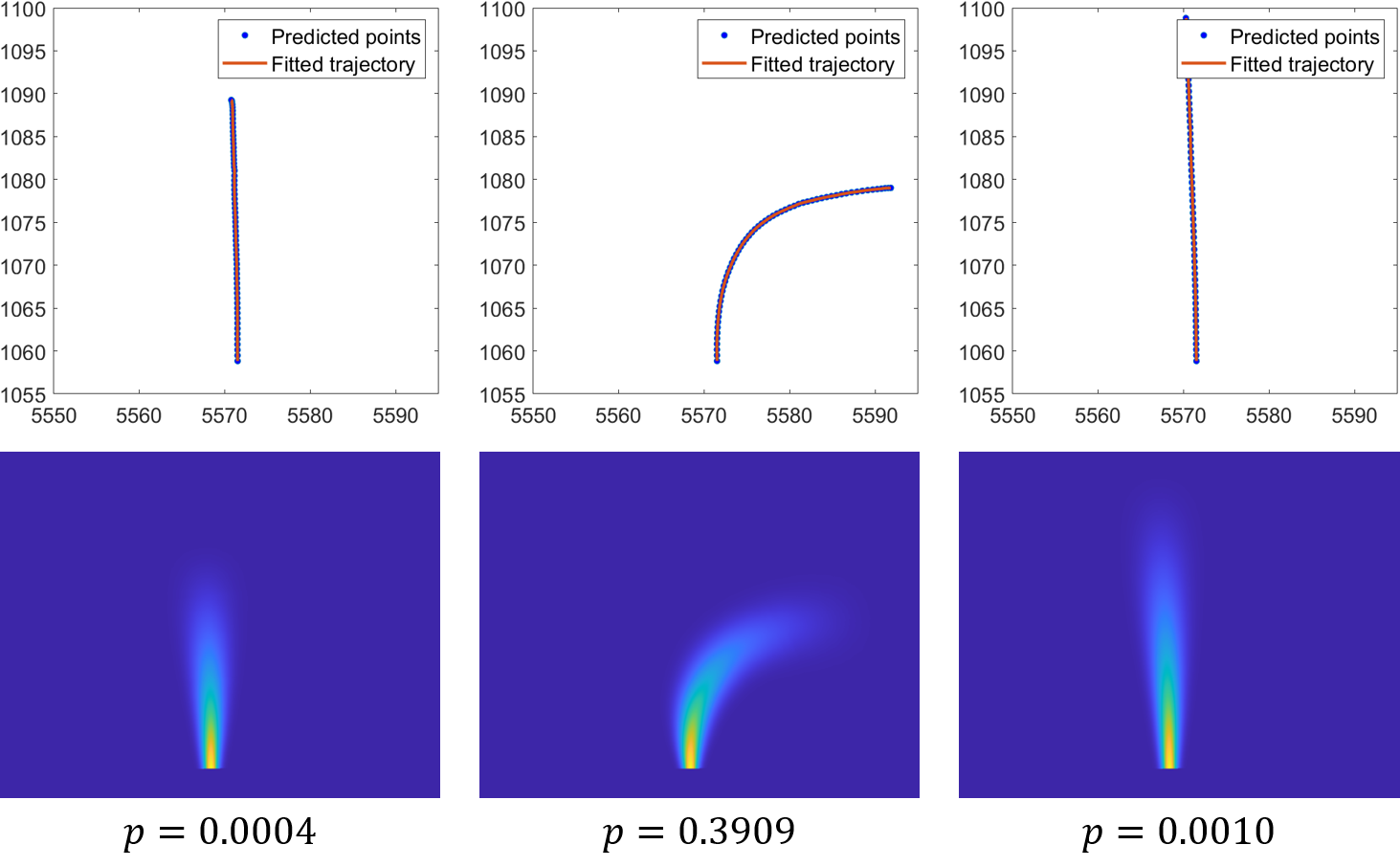} 
  \end{minipage}
  \caption{The $DRP$ distributions for each predicted trajectory (with the probability). They are combined together to obtain the $EDRF$.}
  \label{fig:fitted_traj_and_DRP}
\end{figure}

Fig.~\ref{fig:frenet} and Fig.~\ref{fig:torus} illustrate the establishment of the Frenet coordinate system and the evolution of the torus cross-section when the predicted trajectory follows a sinusoidal curve. This example clearly demonstrates the advantages of prediction-based trajectory methods over kinematic-based methods (Fig.~\ref{fig:prediction-good}). Furthermore, to provide a visual understanding of the modeling results, we take the scenario corresponding to Fig.~\ref{fig:traj_pred_exams}(b) as an example, and the $DRP$ distributions for each predicted trajectory (with the probability) are depicted in Fig~\ref{fig:fitted_traj_and_DRP}. The parameters to model $EDRF$ are shown in Table~\ref{tab:DRP_param} (according to \cite{kolekar2020human, wang2016driving}).

% TODO: 图2b里面的六条轨迹，画出对应的DRP。

\begin{table}[h]
    \centering
    \caption{Parameters in the EDRF Model}
    \begin{tabular}{ccccccc}  % 四列，全部居中
        \toprule  % 顶部粗线
        $q$ & $b$ & $k$ & $c$ & $\alpha$ & $\beta$ & $\gamma$ \\  % 表头
        \midrule  % 中间线
        0.0001 & 0.04 & 1 & 0.5 & $1.566\times 10^{-14}$ & 6.687 & 0.3345\\  % 数据行
        \bottomrule  % 底部粗线
    \end{tabular}
    \label{tab:DRP_param}
\end{table}

\subsection{The complete $EDRF$ model}

In multimodal trajectory prediction, the complete $DRP$ is the weighted sum of the distributions corresponding to each individual trajectory, that is:

\begin{equation}
DRP^ C(x,y)=\sum_{i=1}^n p_i \cdot a(s_i) \cdot \exp{(-\frac{d_{i}^2}{2\sigma(s_i)^2})}
\end{equation}

where $p_i$ represents the probability associated with the $i$-th trajectory, while $s_i$ and $d_i$ denote the coordinates of point $(x,y)$ in the Frenet coordinate system corresponding to the $i$-th trajectory. 

$DRP$ can be seen as a model representing the probability of an event occurring. Modeling the driving risk additionally requires accounting for the consequences of that event. Generally, the consequences of an event are directly related to parameters such as the vehicle's mass, type, and velocity. According to \cite{wang2015driving2}, the consequence can be modelled by virtual mass of the vehicle, which is define as:

\begin{equation}
M=m \cdot T \cdot (\alpha \cdot v^{\beta}+\gamma)
\end{equation}

Therefore, the complete $EDRF$ model is the product of the $DRP$ and the virtual mass $M$:

\begin{equation}
EDRF(x,y)=DRP^C(x,y) \cdot M
\end{equation}

\begin{figure}[htbp]
  \centering
  \begin{minipage}{1\linewidth}
    \centering
    \includegraphics[width=0.7\linewidth]{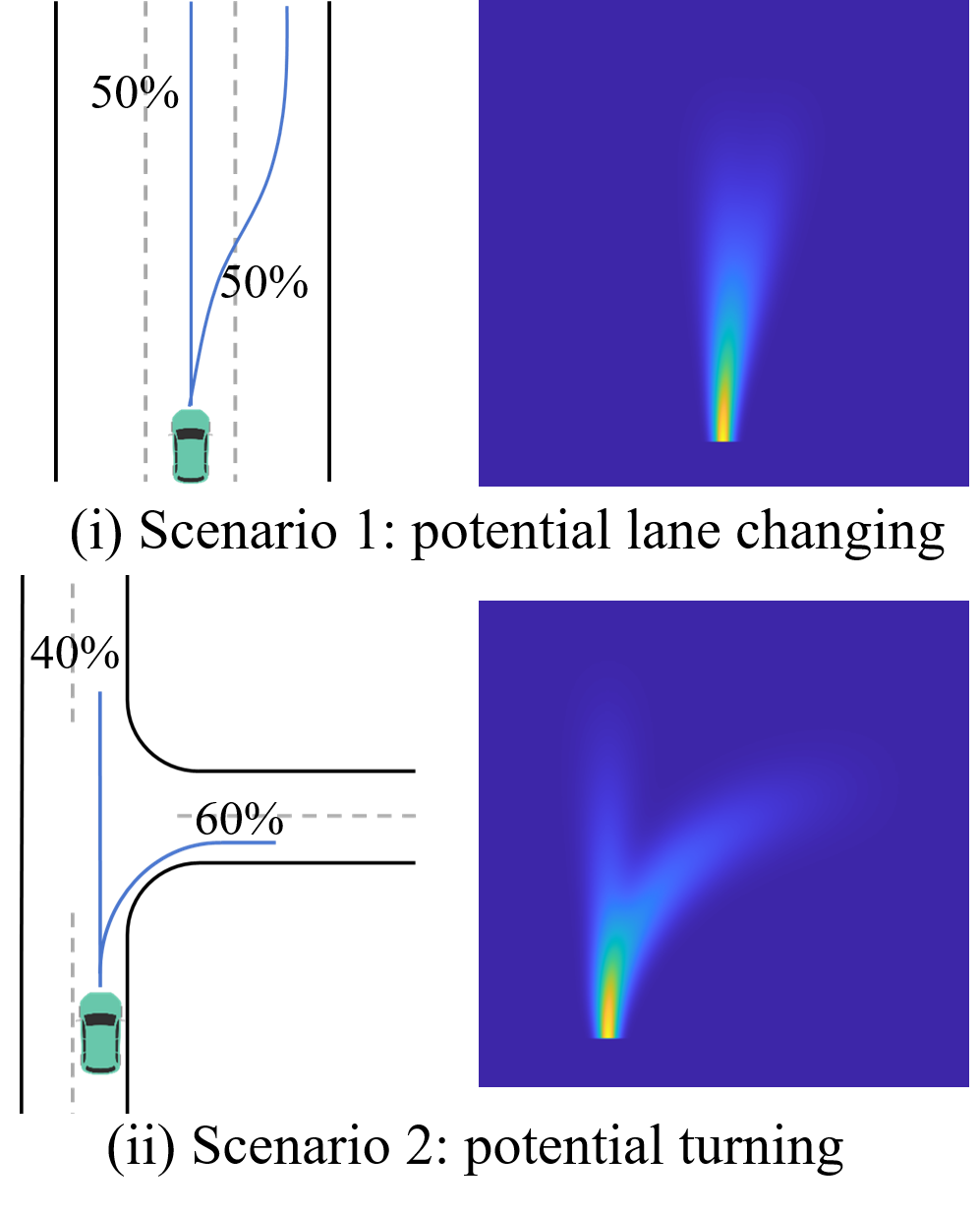} 
  \end{minipage}
  \caption{The $EDRF$ for potential lane changing and potential turning scenarios.}
  \label{fig:DRPC_sample_scenarios}
\end{figure}

\begin{figure}[htbp]
  \centering
  \begin{minipage}{1\linewidth}
    \centering
    \includegraphics[width=0.8\linewidth]{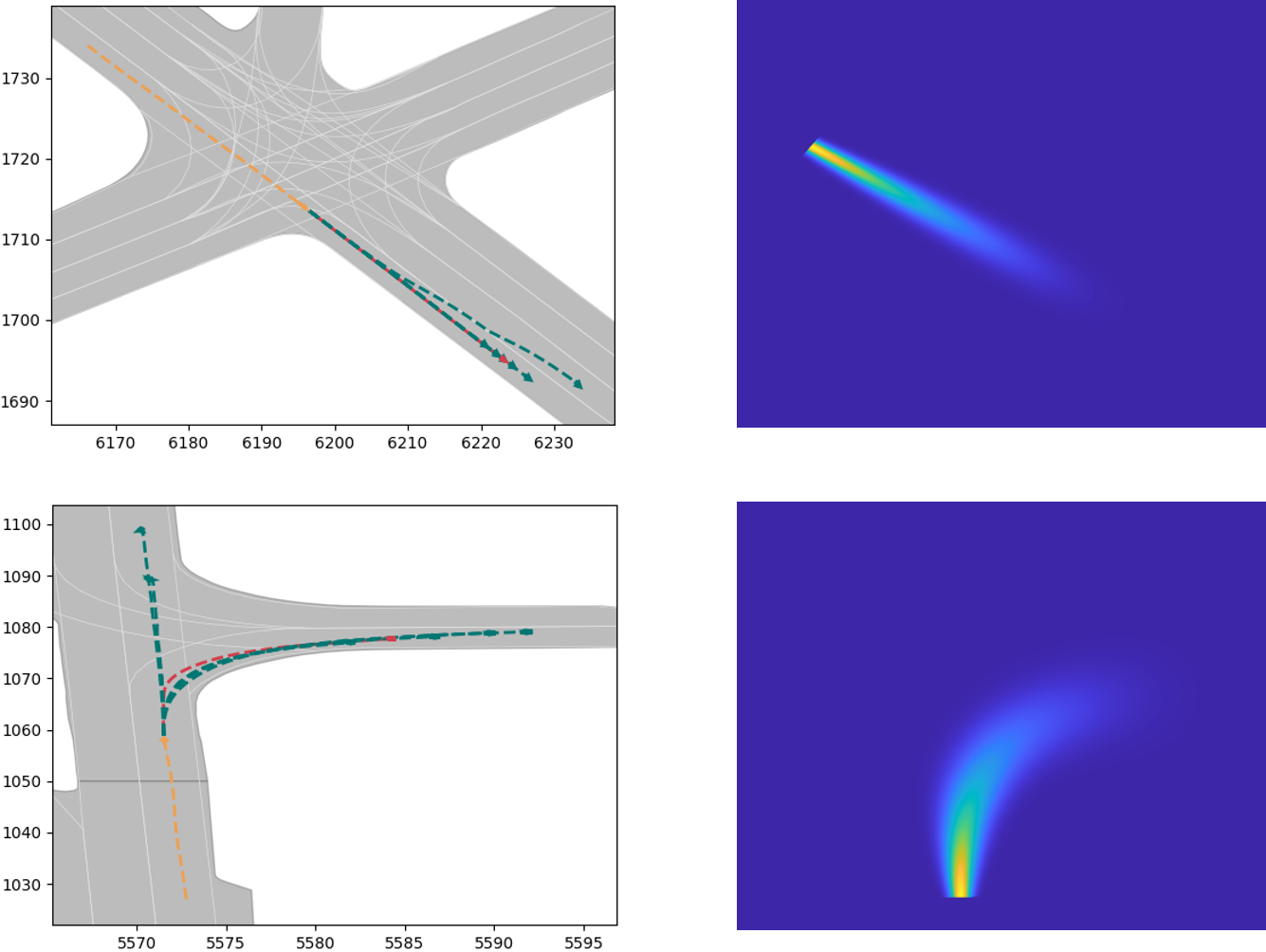} 
  \end{minipage}
  \caption{The $EDRF$ for scenarios in the dataset.}
  \label{fig:DRPC_dataset_scenarios}
\end{figure}

In summary, the $EDRF$ is parameterized by $q, b, k, c, \alpha, \beta, \gamma$ (with their respective numerical values detailed in Table~\ref{tab:DRP_param}), and is determined by the results of multimodal trajectory prediction as well as ego-vehicle state parameters such as mass, type and velocity. To evaluate the effectiveness of the model, we first defined two scenarios: potential lane changing and potential turning. The $EDRF$ for these scenarios are shown in Fig~\ref{fig:DRPC_sample_scenarios}. Subsequently, scenes from the dataset are used. We calculate the $EDRF$ in scenarios (a) and (b) of Fig~\ref{fig:traj_pred_exams} and depict them in Fig~\ref{fig:DRPC_dataset_scenarios}. The results demonstrate that the $EDRF$ is capable of representing the uncertainty in driving intentions, as well as the propagation of this uncertainty along the predicted trajectory.

% TODO:TABLE2

\section{Applications Based on The $EDRF$} \label{Sec:Applications}

In this section, we will introduce applications based on the $EDRF$, which is utilized for three distinct tasks: traffic risk monitoring, ego-vehicle risk analysis, and motion and trajectory planning.

\subsection{Traffic Risk Monitoring}
In the traffic environment, risk is composed of interactions among various entities. The vast majority of traffic accidents can be decomposed into conflicts between two traffic entities. Therefore, we primarily focus on the interaction risk between pairs of entities. For any two traffic participants $i$ and $j$, we first establish the $EDRF$ for each, then the interaction risk between them, denoted as $IR_{ij}$, is defined as follows:

\begin{equation}
IR_{ij}(x,y)=EDRF_i(x,y) \cdot EDRF_j(x,y)
\label{eqn:IR}
\end{equation}

This implies that interaction risk exists only where the $EDRF$ of both entities is non-zero. Additionally, for location $(x,y)$ where interaction risk is present, its calculation can be decomposed into $IR_{ij}(x,y)=DPR_i^c(x,y)\cdot DPR_j^c(x,y) \cdot M_i \cdot M_j$. Here, $DPR_i^c(x,y)\cdot DPR_j^c(x,y)$ quantifies the probability of both entities being simultaneously present at $(x,y)$, and thus colliding, while $M_i \cdot M_j$ quantifies the severity of the collision. Finally, we determine the level of risk $F_{ij}$ between entities $i$ and $j$ by taking the maximum of the interaction risk at all positions:

\begin{equation}
F_{ij}(x,y)=\max_{x,y} IR_{ij}(x,y)
\label{eqn:F}
\end{equation}

\begin{figure}[htbp]
  \centering
  \begin{minipage}{1\linewidth}
    \centering
    \includegraphics[width=0.95\linewidth]{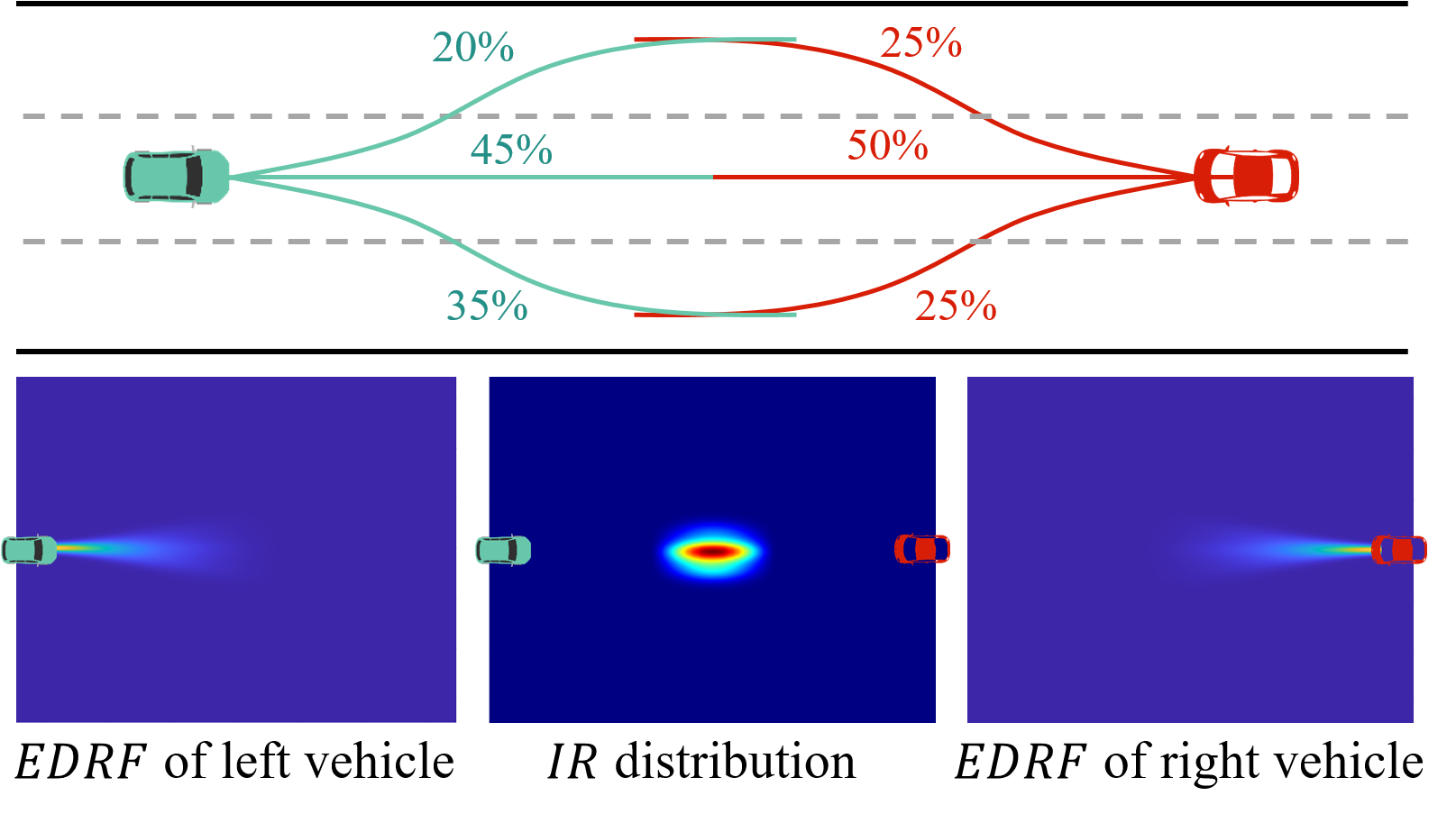} 
  \end{minipage}
  \caption{Application on traffic risk monitoring: the designed head-on scenario, the respective $EDRF$ of the two vehicles and the distribution of their interactive risk ($IR$).}
  \label{fig:app1}
\end{figure}

The scenario of two vehicles traveling head-on is utilized as an example. Fig~\ref{fig:app1} illustrates the respective $EDRF$ of the two as well as the distribution of their interaction risk ($IR$). In practical applications, a threshold $F_{thld}$ can be set for the risk $F$. if the risk intensity between two entities in the traffic environment exceeds $F_{thld}$, it may lead to a traffic accident and should call attention to the driver of HDV or intervene in the motion of AV.

\subsection{Ego-vehicle Risk Analysis}
A significant difference between ego-vehicle risk analysis and traffic risk monitoring is that the uncertainty associated with the ego-vehicle is substantially lower than that of predicted trajectories of other vehicles. Moreover, ego-vehicle trajectory prediction does not require multimodal approaches. Therefore, a kinematic-based can be employed for ego-vehicle trajectory prediction: assuming that the vehicle will continue to move according to its current state (position, orientation, steering angle and velocity) over a look-ahead time period $t_{la}$ (6 seconds in this paper).

\begin{figure}[htbp]
  \centering
  \begin{minipage}{1\linewidth}
    \centering
    \includegraphics[width=1.0\linewidth]{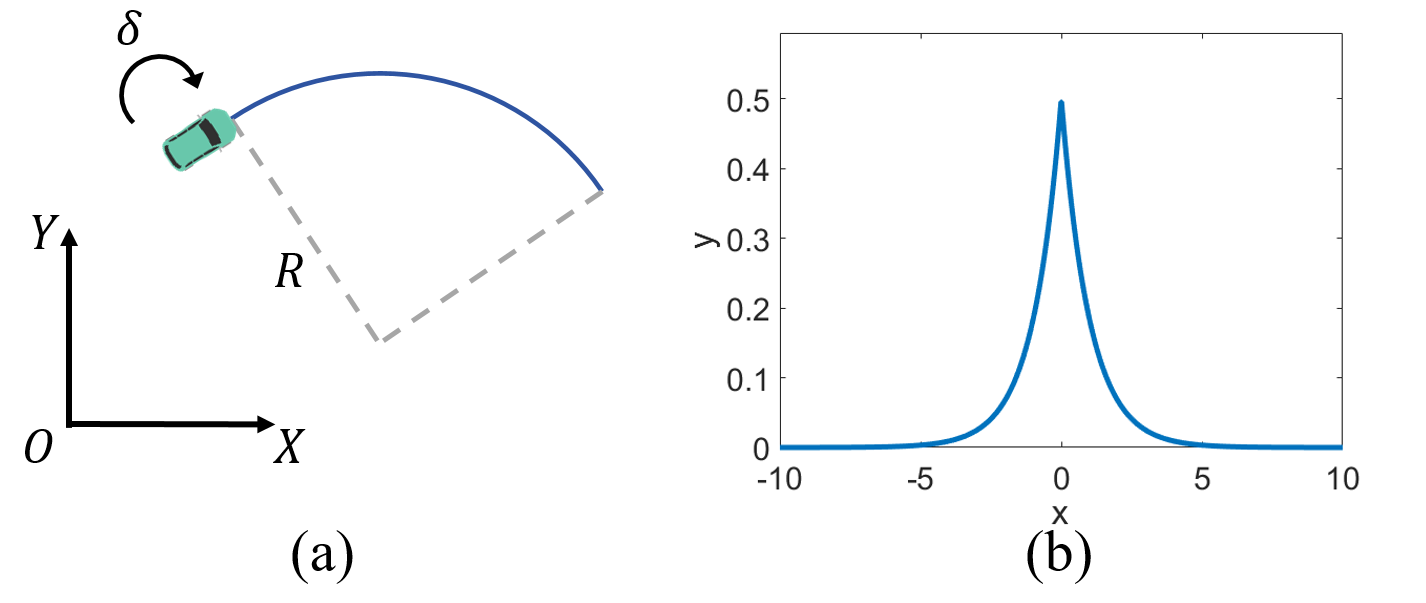} 
  \end{minipage}
  \caption{Kinematic-based future trajectory prediction and the Laplace distribution.}
  \label{fig:kinematic}
\end{figure}

According to the simplified bicycle model, the turning radius of the vehicle can be calculated as:

\begin{equation}
R=\frac{L}{\tan(\delta)}
\end{equation}

where $L$ represents the wheelbase and $\delta$ is the steering angle. Then, combined with other state parameters of ego-vehicle, the future trajectory can be determined (Fig~\ref{fig:kinematic}(a)). 

Furthermore, due to the lower uncertainty of ego-vehicle, using a Gaussian distribution to quantify is not appropriate. Hence, we replace the Gaussian distribution with a Laplace-like one (Fig~\ref{fig:kinematic}(b)) to represent the cross-section of the torus. The height and width of the Laplace-like distribution are also functions of the predicted trajectory s.

\begin{equation}
DPR_{ego}(x,y)=a_{ego}(s) \cdot \exp{(-\frac{|d|}{\lambda(s)})}
\end{equation}

The height $a_{ego}$ also decreases along $s$, but is modeled as a linear function for the lower uncertainty. The trend in width $\lambda$ variation follows the same pattern as equation (\ref{eqn:sigma}), that is:

\begin{equation}
a_{ego}(s)=q_{ego}|s-v \cdot t_{la}|
\end{equation}

\begin{equation}
\lambda(s)=(b_{ego}+k_{ego}\cdot |\delta|)s+c_{ego}
\end{equation}

\begin{figure}[htbp]
  \centering
  \begin{minipage}{1\linewidth}
    \centering
    \includegraphics[width=0.95\linewidth]{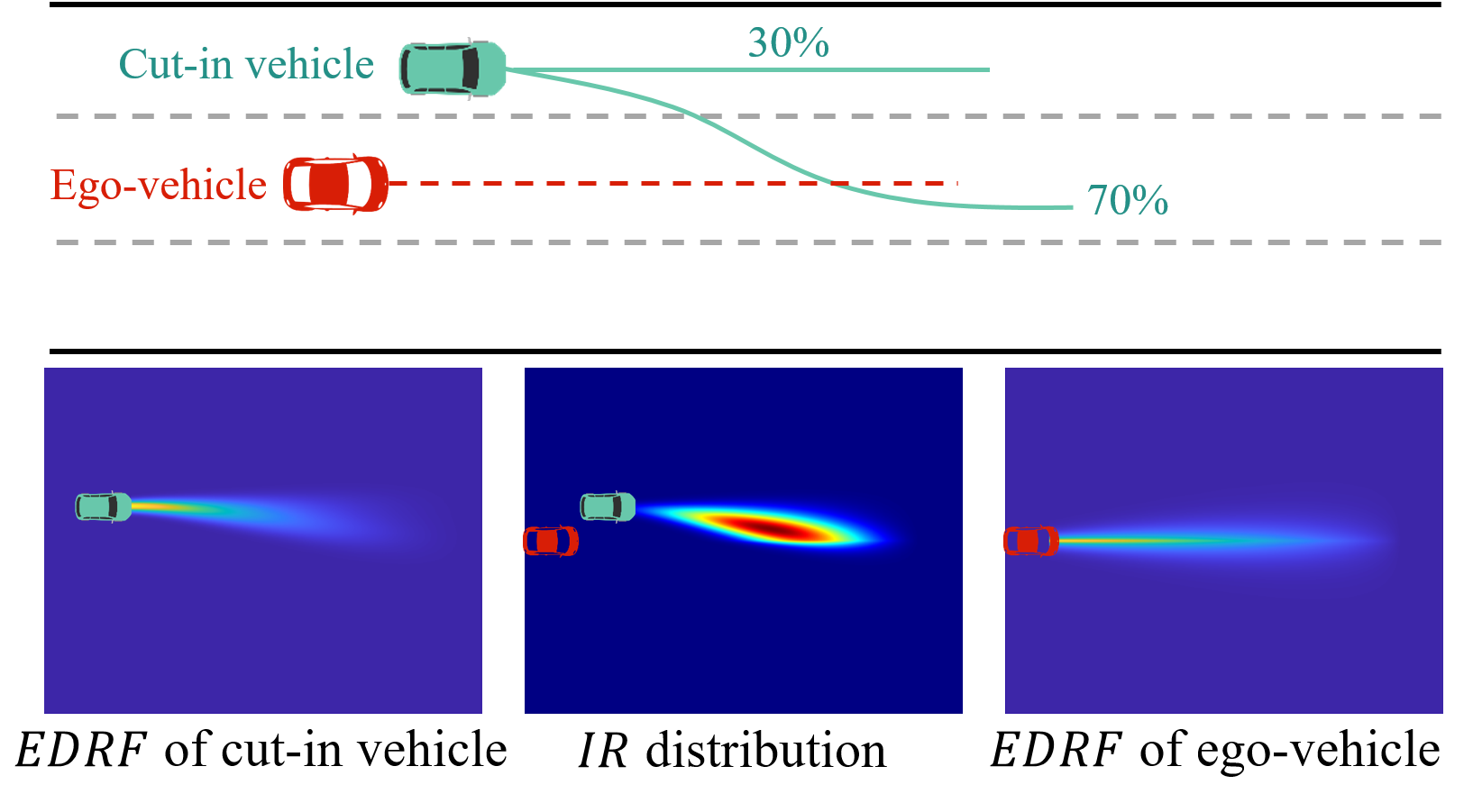} 
  \end{minipage}
  \caption{Application on ego-vehicle risk analysis: the designed potential cut-in scenario, the respective $EDRF$ of the adjacent vehicle and the ego-vehicle, and the distribution of their interactive risk ($IR$)}
  \label{fig:app2}
\end{figure}

The parameters are set as Table~\ref{tab:ego_param}. The calculation of the interaction risk also adheres to equation (\ref{eqn:IR}) and (\ref{eqn:F}). Here we take the potential cut-in of adjacent vehicle as an example scenario. Fig~\ref{fig:app2} shows the $EDRF$ for both the ego-vehicle and the cut-in vehicle , as well as the $IR$ distribution between them. A threshold $F_{thld}$ can similarly be set as a criterion for analyzing the risk situation of ego-vehicle.

\begin{table}[h]
    \centering
    \setlength{\tabcolsep}{17pt}
    \caption{Parameters in the $EDRF$ Model of Ego-vehicle}
    \begin{tabular}{cccc}  % 四列，全部居中
        \toprule  % 顶部粗线
        $q_{ego}$ & $b_{ego}$ & $k_{ego}$ & $c_{ego}$  \\  % 表头
        \midrule  % 中间线
        0.004 & 0.05 & 1 & 0.5 \\  % 数据行
        \bottomrule  % 底部粗线
    \end{tabular}
    \label{tab:ego_param}
\end{table}

\subsection{Motion and Trajectory Planning}

For motion and trajectory planning of AVs, the $EDRF$ and $IR$ can be used in various manners, depending on the specific framework of the planning module. The most straightforward method to integrate the $EDRF$ and $IR$ is to use them as safety assessment metrics for trajectories, thereby guiding the planning module to select safer trajectories. 

For instance, the $EDRF$ and $IR$ can be combined with a trajectory sampling framework, calculating the $EDRF_{ego}$ for each sampled trajectory. Here, as trajectories have already been sampled, there is no need to compute future trajectories based on the current state. Additionally, due to the low uncertainty associated with the ego vehicle, a Laplace-like distribution is still employed for the cross-section of the torus. Subsequently, the $IR$ and $F$ for each trajectory, based on the $EDRF$ of surrounding vehicles, can be determined. These metrics serve as safety indicators for feasible trajectories, which, along with other criteria such as comfort, efficiency, and energy conservation, allow for the selection of the optimal trajectory according to customized standards. 

\begin{figure}[htbp]
  \centering
  \begin{minipage}{1\linewidth}
    \centering
    \includegraphics[width=0.95\linewidth]{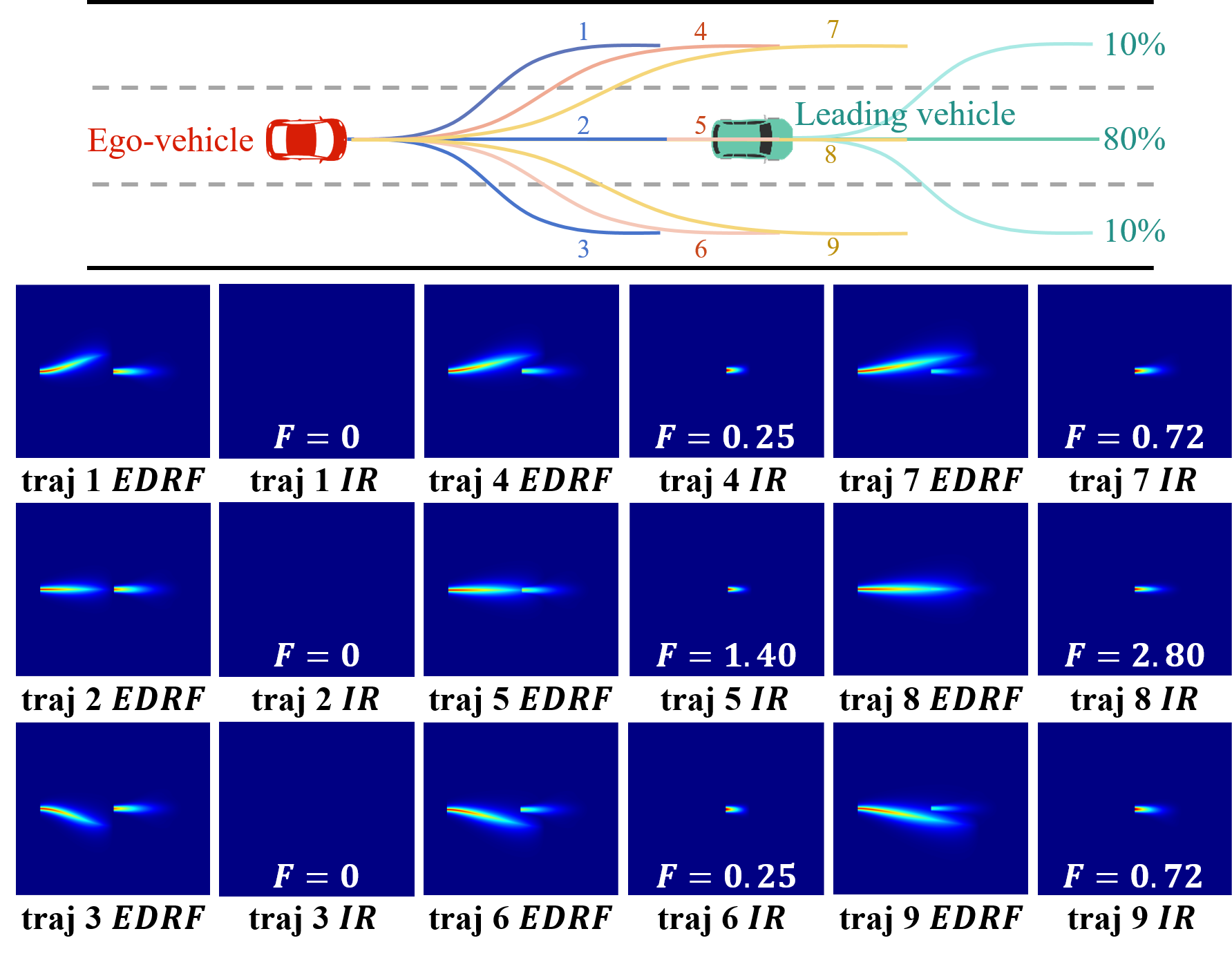} 
  \end{minipage}
  \caption{Application on trajectory planning: the designed scenario where a leading vehicle decelerates, the respective $EDRF$, $IR$ and $F$ of each feasible trajectories.}
  \label{fig:app3}
\end{figure}

We design a scenario where a leading vehicle decelerates as an example, in which different planning methods may yield trajectories for either following with deceleration or overtaking by lane changing. We sample nine feasible trajectories in front of the vehicle, representing options for left lane change / following / right lane change, and acceleration / maintaining velocity / deceleration. Figure~\ref{fig:app3} shows the $IR$ distribution for each of these trajectories. From the safety perspective, deceleration is the best choice; however, when combined with other criteria and their respective weights, different outcomes, such as overtaking by lane changing, may also be recommended.

\section{Conclusion} \label{Sec:Conclusion}

In this paper, the Enhanced Driving Risk Field ($EDRF$) for traffic risk assessment is proposed. By integrating multimodal trajectory prediction results with Gaussian distribution models, the driving risk probability ($DRP$) model is presented, where the uncertainty of drivers’ behavior is quantitatively captured. The $EDRF$ of traffic entities is modeled as the product of the complete $DRP$, which represents the probability of an event occurring, and virtual mass, which stands for the consequence of the event. We also propose the applications based on the $EDRF$. The interaction risk ($IR$) between any two entities is defined, upon which the $EDRF$ is applied across various tasks (traffic risk monitoring, ego-vehicle risk analysis and motion and trajectory planning) using a unified manner. Example scenarios for each application are provided, and the results demonstrate the effectiveness of the model.

Currently, the effectiveness of the $EDRF$ has only been preliminarily validated. In the future, we plan to conduct large-scale validation of the $EDRF$ model across various applications by using publicly open-source datasets or by collecting real-world or driving simulator data. Moreover, field theory holds significant potential to generate human-like driving behaviors for AVs. Therefore, future studies will also explore the relationship between the $EDRF$ and human-like driving behaviors.

\bibliographystyle{IEEEtran}
\bibliography{mybibfile}

\end{document}